%% file: main.tex
\documentclass[10pt,twocolumn,letterpaper]{article}

 \usepackage{cvpr}              %

\definecolor{cvprblue}{rgb}{0.21,0.49,0.74}
\usepackage[pagebackref,breaklinks,colorlinks,allcolors=cvprblue]{hyperref}

\input{sec/_preamble}

\input{sec/_math}

\def\paperID{xxx} 
\def\confName{CVPR}
\def\confYear{2025}

\title{It Takes Two: Real-time Co-Speech Two-person's  Interaction Generation via Reactive Auto-regressive Diffusion Model}

\author{Mingyi Shi$^{1\ast}$,
Dafei Qin$^{1\ast}$,
Leo Ho$^{1,2}$,
Zhouyingcheng Liao$^{1}$,
Yinghao Huang$^{3}$
\\
Junichi Yamagishi$^{4}$,
Taku Komura$^{1,2}$
\vspace{0.2cm}
\\
{\normalsize $^{1}$ The University of Hong Kong \quad $^{2}$ Centre for Transformative Garment Production} \\ 
{\normalsize $^{3}$ Great Bay University \quad $^{4}$ National Institute of Informatics, Tokyo  }    \\
{\normalsize {$^{\ast}$ Equal Contributions.}}
}

\begin{document}
\maketitle

\let\thefootnote\relax\footnotetext{arXiv preprint, 2024}

\begin{strip}\centering
\includegraphics[width=\textwidth]{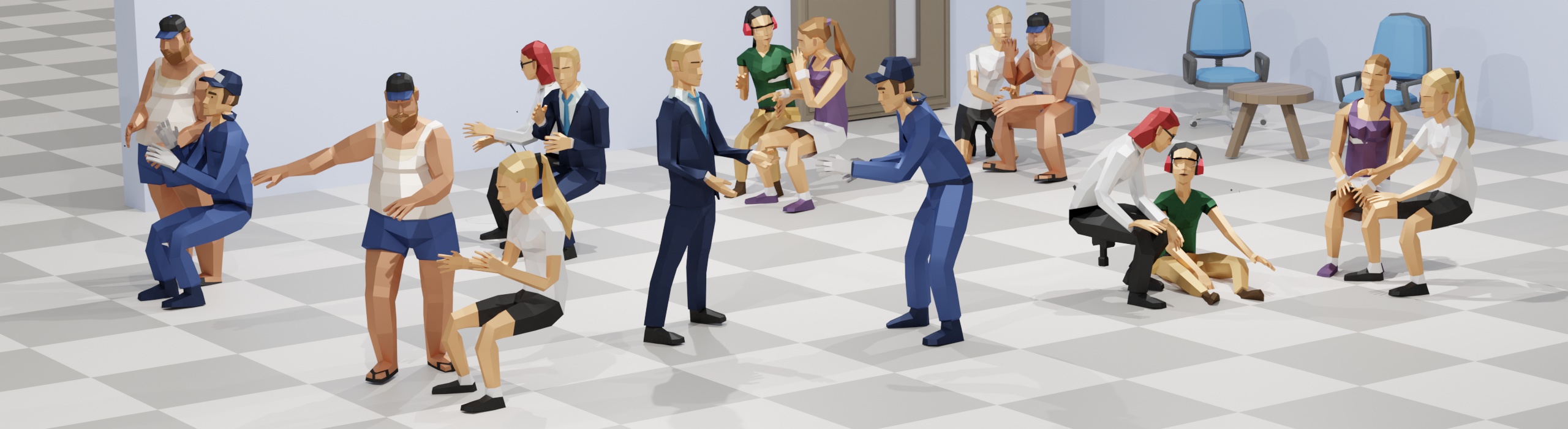}
\captionof{figure}{Our system addresses a novel task, that takes the speech of two persons as input to generate dynamic full-body interactions in real-time. To achieve this, we designed an audio-driven, auto-regressive diffusion model that generates two-person motion, with the guidance of motion trajectory to improve controllability. To enrich the diversity of these interactions, we captured a new dataset that includes a wide range of daily conversational scenarios, and short-order execution.}
\label{fig:teaser}
\end{strip}

\input{sec/0_abstract}    
\input{sec/1_intro}

\input{sec/2_related}

\input{sec/3_method}

\input{sec/4_experiments}

\input{sec/5_conclusion}

{
    \small
    \bibliographystyle{ieeenat_fullname}
    \bibliography{main.bib}
}

 \input{sec/X_suppl}

\end{document}

%% file: sec/_preamble.tex
\usepackage{graphicx}
\usepackage{amsfonts}
\usepackage{amsmath}
\usepackage{booktabs}
\usepackage{xcolor}
\usepackage{multirow}
\usepackage[normalem]{ulem}
\usepackage[super]{nth}
\usepackage{colortbl}
\usepackage{xspace}
\usepackage{cuted}
\usepackage{listings}

\newif\ifdraft
  \draftfalse
\ifdraft
\newcommand{\myc}[1]{{\color{blue}\textbf{Mingyi:} #1}}
\newcommand{\tkc}[1]{{\color{cyan}\textbf{Taku:} #1}}
\newcommand{\dfc}[1]{{\color{orange}\textbf{Dafei:} #1}}
\newcommand{\lhc}[1]{{\color{green}\textbf{Leo:} #1}}
\newcommand{\my}[1]{{\color{blue}#1}}
\newcommand{\tk}[1]{{\color{cyan}#1}}
\newcommand{\df}[1]{{\color{orange}#1}}

\else
\newcommand{\myc}[1]{}
\newcommand{\tkc}[1]{}
\newcommand{\dfc}[1]{}
\newcommand{\lhc}[1]{}
\newcommand{\my}[1]{{\color{black}#1}}
\newcommand{\tk}[1]{{\color{black}#1}}
\newcommand{\df}[1]{{\color{black}#1}}

\fi

\usepackage{xspace}

\newcommand{\keep}[1]{}
\newcommand{\old}[1]{}

\usepackage{url}
\usepackage{hyperref}

%% file: sec/_math.tex
\DeclareMathOperator*{\mse}{mse}

\newcommand{\sample}{ \mathbf{x} }
\newcommand{\samplepred}{ \hat{\mathbf{x}} }
\newcommand{\net}{ \mathcal{G} }

\newcommand{\loss}{ \mathcal{L} }

%% file: sec/0_abstract.tex
\begin{abstract}

Conversational scenarios are common in the real-world, yet existing co-speech motion synthesis approaches often fall short in these contexts, where one person's audio and gestures will influence the other's responses. 
Additionally, most existing methods rely on offline sequence-to-sequence frameworks, which are unsuitable for online applications. 
In this work, we introduce an audio-driven, auto-regressive system designed to synthesize dynamic movements for two characters during a conversation. 
At the core of our approach is a diffusion-based full-body motion synthesis model, which is conditioned on the past states of both characters, speech audio, and a task-oriented motion trajectory input, allowing for flexible spatial control. 
To enhance the model's ability to learn diverse interactions, we have enriched existing two-person conversational motion datasets with more dynamic and interactive motions. 
\my{We evaluate our system through multiple experiments to show it outperforms across a variety of tasks, including single and two-person co-speech motion generation, as well as interactive motion generation.}
To the best of our knowledge, this is the first online system capable of generating interactive full-body motions for two characters from speech.

\end{abstract}

%% file: sec/1_intro.tex
\section{Introduction}
\label{sec:intro}
Studying nonverbal communication, including body gestures, has captivated researchers across psychology, natural language processing, machine learning, and computer graphics. The intricate full-body motion during conversations carries significant implications, conveying emotions, indicating positions, emphasizing agreement, and enhancing communication between speakers. Developing a system capable of replicating such behavior is not only of practical interest but also offers insights into how humans orchestrate their movements, bridging the realms of application and scientific inquiry.

With the recent progress in the digital co-speech dataset, the task of synthesizing body gestures from speech has gained more attention. Existing methods have shown promising results in generating plausible body gestures from speech~\cite{pang2023bodyformer, alexanderson2023listen, ahuja2020style, moglow}. 
However, these methods often overlook the interactive nature of human conversations, where the body gestures of one speaker are influenced by the speech and motion of the other speaker. 
Moreover, most models assume the characters are standing still and simply moving the upper part of the body in sync with the speech. The limited dynamics in the existing dataset ~\cite{cassell2001beat} is one reason, more importantly, their models do not take into account of the body global movements, which are crucial for modeling the interaction.
Although some recent diffusion models~\cite{gesturediffuclip, alexanderson2023listen,arfriend} can produce characters to wander around while talking, control is applied either through speech or language prompts only. Although such models allow the style of how a character talks to be controlled, they cannot produce synchronized interactions such as handshaking, passing one object to another, or any other realistic physical interactions that we observe in daily life, which are strongly conditioned on the other person's motion. Another obstacle that makes the effective modeling of realistic and expressive interaction behaviors difficult is that these interaction behaviors are still lacking in existing datasets. Most existing datasets feature two persons talking with each other and making gestures while standing still \cite{arfriend, ng2024audio2photoreal}, without interesting interaction patterns like passing one object to another, or two persons walking to the same destination side by side. 

The setup of most existing data-driven techniques inhibits their usage for real-world applications. 
For one thing, techniques based on temporal convolution~\cite{ahuja2020style}, normalizing flows~\cite{moglow}, transformers~\cite{pang2023bodyformer} and diffusion models~\cite{alexanderson2023listen, ng2024audio2photoreal}, are sequence-to-sequence models that require the entire speech to be given in advance, making them difficult to be applied for online applications. 
For another,  humans often perform gestures and speech at the same time, in response to the contents of the previous speech, the body gesture and facial expression of the other person, and the current physical and psychological state of the person; for example, the body gestures would be different according to the relative distance between the speakers, whether they are standing still or walking, etc. These additional conditions are not considered by previous methods~\cite{ao2023gesturediffuclip, liu2024emage}.

To formulate conversational interaction, we introduce a novel task that utilizes the speech of two individuals to generate dynamic full-body motion. We propose an online co-speech motion generation system for two persons, capable of producing the full-body motion of two characters in sync with their speech and past status. The system captures both the high-level spatial context and the low-level joint animations of the characters to produce realistic and dynamic interactions.
To enhance the diversity of interactive motion, we collected a new dataset featuring the motions of two individuals across various scenarios. The system is trained on this dataset, alongside a single-person co-speech dataset, employing a random masking strategy inspired by classifier-free guidance in diffusion models.
Our system runs in an autoregressive manner, which is closer to the real-world scenarios where people condition their body motion on their current talking context and the other person's audio and motion feedback, and thus can be applied for real-time character controllers for more expressive and interactive motion synthesis.  
From the experiments, our system outperforms other methods in generating two-person interactive motion, producing better contextual alignment with the speech, and the motion of the other person. We also conduct a comprehensive evaluation of the system to show the superiority of our autoregressive system in real-time applications.

\begin{figure}[t]
	\centering
	\includegraphics[width=1\columnwidth]{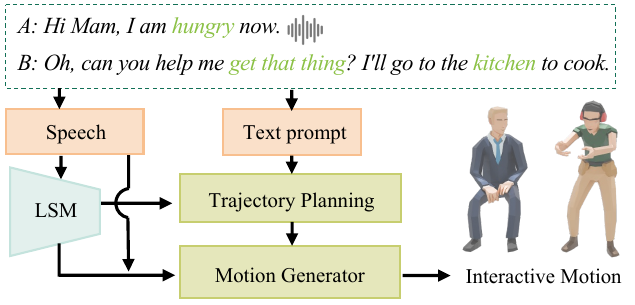}
	\caption{\textbf{Concept diagram.} \my{Our system obtains two persons' speech as input to generate full-body motion. We employ Large-Speech-Model (LSM) to extract the semantic token, which are then fed into our autoregressive motion generation module, to produce interactive motion with the guidance of predicted trajectory.}
    }
	\label{fig:system}
\end{figure}

To the best of our knowledge, this is the first work that can autoregressively generate the full body motion of two characters in response to their speech and the motion of the other character.
Consequently, our system can be applied for various online applications such as VR systems, computer games and customer representatives. Both the source code and the dataset will be made publicly accessible online.
In summary, the contribution of our system includes:
\begin{itemize}
    \item A novel formulation of the task of generating full-body motion of two characters during their conversation.
    \item \my{A multi-conditional motion generation module that can interactively generate realistic movements in response to speech, body motion, and 2D body root trajectories.}
    \item \my{Our online framework achieves state-of-the-art performance in co-speech motion generation tasks, for both single and two-person scenarios.}
    \item An enriched dataset comprising diverse interaction.

\end{itemize}

%% file: sec/2_related.tex
\section{Related Work}

\paragraph{Co-speech Gesture Synthesis}
Human motion and speech are closely interrelated, and their interaction contributes to effective communication, as demonstrated by numerous cognitive science literatures~\cite{cassell1999speech,de2012interplay,wagner2014gesture}.
Rule-based methods~\cite{kopp2006towards, cassell2001beat,huang2012robot} synthesize motion sequences from speeches based on pre-constructed knowledge bases. 
Data-driven are proposed to reduce the manual labor of rules creation. \citet{neff2008gesture} applies a statistical model to synthesize gestures and body motion accompanying speech.
Gesture controllers~\cite{levine2010gesture} designs an inference layer based on a Conditional Random Field to map vocal prosody features to a hidden representation of gesture kinematics.
Recent works use powerful deep neural networks for end-to-end co-speech gesture synthesis. Early attempts use deterministic models like MLPs \cite{kucherenko2020gesticulator}, CNNs \cite{habibie2021learning}, and RNNs \cite{hasegawa2018evaluation, yoon2020speech, liu2022learning} to learn the mapping between speech and gestures. Later, generative models are leveraged to better learn the many-to-many mapping between the speech and the gestures. These include 
GAN-based methods \cite{habibie2021learning,bhattacharya2021speech2affectivegestures}, VAEs \cite{li2021audio2gestures, ao2022rhythmic, pang2023bodyformer,xiao2024eggesture}, normalizing flow~\cite{alexanderson2020style} and diffusions ~\cite{deichler2023diffusion, alexanderson2023listen, ao2023gesturediffuclip, liu2024emage, Chhatre_2024_CVPR}. \citet{nyatsanga2023comprehensive} provides a comprehensive review of these methods. 
\tk{These methods generate gestures based solely on an individual’s speech and ignore crucial context provided by a conversational partner \cite{kucherenko2023genea,kucherenko2024evaluating}. Also, most of the methods model only local gestures \cite{yoon2020speech, ao2022rhythmic, Chhatre_2024_CVPR} without addressing the global position of the speaker. As a result, these approaches generate gestures for a stationary figure, failing to adapt to interactive conversations.}

\tk{
The closest one related to us is~\cite{ng2024audio2photoreal}, which takes two-person audio as input and generates full-body motions. 
Despite their method produces realistic motion of the face and body given the speech, the two characters simply stands facing each other with little motion or interaction. 
}
On the contrary, we generate co-speech gestures by additionally conditioning on the speech and gestures of the partner, supporting more immersive, context-aware gesture synthesis.  Besides, our approach incorporates dynamic global positioning with an independent trajectory prediction module, enabling speakers to interact naturally within the conversation space. 
\paragraph{Multi-character Interactions Modeling}
In most scenes where two or more characters are present, modeling their interactions is crucial for creating realistic and immersive content.
Given a dancer's motion, \citet{hsu2004example} generates the partner motion by retrieving it in databases.
Liu et al.~\citet{liu2006composition} employ a spacetime optimization to create multi-character motions from short single-character motions.
Kwon et al.~\citet{kwon2008two} adopt a dynamic Bayesian network to train a motion transition model to guide the generation of two-character motion streams.
Kim et al.~\citet{kim2009synchronized} utilize Laplacian motion editing to allow for multi-character motion editing.
Another line of research~\cite{shum2008interaction, kim2012tiling, won2014generating} utilizes the Interaction Patch, an extension of the Motion Patch~\cite{lee2006motion}, to generate interactions between multiple characters.

Advancements in deep learning have further enhanced the generation of multi-character interactions. 
\citet{kundu2020cross} present a cross-conditioned recurrent neural network to synthesize inter-person motion interactions.
\citet{starke2020local} propose a local motion phase that learns asynchronous movements of each bone and its interaction with external objects.
Generative models like GANs ~\cite{men2022gan,goel2022interaction} and diffusions ~\cite{tevet2023human,zhang2022motiondiffuse,chen2023mld,zhou2023emdm,alexanderson2023listen} are adopted for interaction synthesis.
ComMDM~\cite{shafir2024human} learns a communication block to generate two-person motions from text prompts based on a pre-trained motion diffusion model as the generative prior.
InterGen~\cite{liang2024intergen} devises two cooperative transformer-style networks with shared weights and mutual attention that generate diverse two-person motions simultaneously.
MoMat-MoGen~\cite{cai2023digital} adopts an active-passive mechanism to model the two-person interaction via a hybrid method of text-based motion matching and diffusion-based motion generation.
Duolando~\cite{siyao2024duolando} proposes a GPT-based model to predict the dancing motion of a follower given a leader dancer's motion conditioned on the coordinated information of the music. 
ReMoS \cite{ghosh2024remos} and InterMask\cite{javed2024intermask3dhumaninteraction} synthesize the reaction of one person conditioning on the partner's motion. 
FreeMotion \cite{ke2024free} enables the conditioning of an arbitrary number of partner motions. 

These methods generate motion patterns based on textual descriptions and overlook the rich, influential role of audio information, which is integral to capturing the full context and emotional nuance of real-time conversation. A closer related work is ARfriend \cite{arfriend}, which captures the motion and speech of two actors and trains a diffusion model to animate the interaction of two persons. However, their model is trained in a sequence-to-sequence way. Large system latency prevents them from being applied in real-time scenarios.
In contrast, our audio-driven method autoregressively generates two-person interactions in real-time, with the support of explicit trajectory control.

%% file: sec/3_method.tex
\section{Methodology}

In this section, we provide a detailed overview of our system. Given the speech inputs $[\mathbf{S}^A, \mathbf{S}^B]$ from two persons, our system generates the full-body motion outputs $[\mathbf{M}^A, \mathbf{M}^B] \in \mathbb{R}^{N \times (J \times Q + 3 + 2)}$ for two characters, where $N$ represents the number of frames, $J$ denotes the number of joints, $Q$ indicates the number of rotation features, and $\mathbb{R}^3$ accounts for the root global displacements, with an additional $\mathbb{R}^2$ for the foot contact label.
We begin by describing the fundamental data processing steps in Sec.~\ref{sec:input}, which are designed for semantic and reliable control. 
Next, we introduce the key component of our system: a dual-streaming motion generator, detailed in Sec.~\ref{sec:multi-condition}. This module takes speech, trajectory, and the past states of both individuals as inputs to generate two-person's motion simultaneously. 
Finally, in Sec.~\ref{sec:autoregressive}, we explain how the generator is employed in an autoregressive manner to ensure responsiveness and realism in long-term scenarios.

\subsection{Data Processing}
\label{sec:input}

\paragraph{Speech}
\label{par:speech}
Speech contains both acoustic details and semantic information.
For acoustics, we follow previous work~\cite{alexanderson2023listen} to use the \textit{librosa} library~\cite{mcfee2015librosa} to convert the speech signal into Mel-spectrograms, denoted as $\mathbf{s}^{\text{mel}} \in \mathbb{R}^{27}$. To further enhance rhythm alignment, we implement a smoothed 1D curve by multiplying the onset points with their amplitude~\cite{gesturediffuclip}, represented as $\mathbf{s}^{\text{rhy}} \in \mathbb{R}^{1}$.
For semantics, we observed that commonly used text-based embeddings from previous studies~\cite{pang2023bodyformer, gesturediffuclip, emage2024}, such as BERT~\cite{devlin2018bert} and CLIP~\cite{ramesh2022hierarchical}, tend to overcrowd the feature space, reducing alignment between training and test distributions, and consequently impairing generalizations. Therefore, we opted to tokenize the speech data directly using a pre-trained large speech language model~\cite{zhang2023speechtokenizer}, \my{which produces compact and discrete speech tokens, leading to better generalization in our experiments}.  Each 20ms segment of audio will be quantized into tokens with 8 dimensions, from which we extract only the first dimension containing the semantic information, denoted as $\mathbf{s}^{\text{sem}} \in \mathbb{R}^{1}$. 

\paragraph{Trajectory}
Unlike previous co-speech motion generation approaches \df{which produce stand-still talking avatars}, our system aims for more dynamic and interactive motion generation \df{suitable for daily scenarios}. However, this also introduces more uncertainty and complexity, \df{as interactions happen differently in different environments like home, office and restaurant}. 
\tk{
To enhance the reliability and flexibility of the system, 
we synthesize the motion in two stages: first, the root trajectory projected on the ground, and then the full body motion using the root trajectory as an additional control signal. This design provides flexibility to either make the synthesis fully automatic, constrain the root motion of one character, or manually design the trajectory/ies for one or both characters.}     
The trajectory in frame $n$ includes the \df{planner} body positions $\mathbf{T}_n^{\textit{pos}} \in \mathbb{R}^{2}$ and facing directions $\mathbf{T}_n^{\textit{rot}} \in \mathbb{R}^{Q}$ of both individuals.
\tk{
For automatic prediction of the trajectory, 
denoted as $\net_{\text{traj}}$, we utilize the same diffusion model as the full body  
that receives multiple conditions such as speech, activity values and end locations as input.
See Sec.\ref{sec:multi-condition} and supplementary material for the details. 
}

\begin{figure*}[t]
    \centering
    \includegraphics[width=\textwidth]{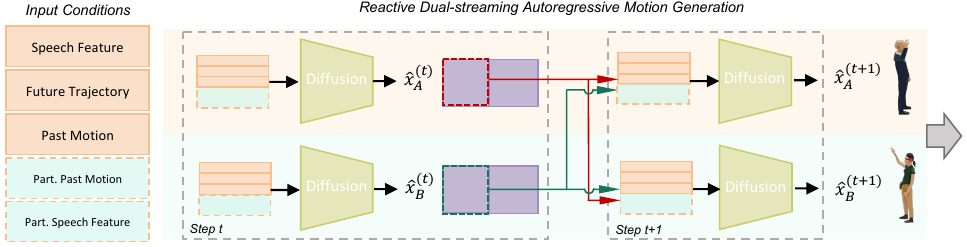}
    \caption{The overview of our autoregressive motion generator. 
    Through a dual streaming design, the motion of two persons are generated simultaneously. 
    For each prediction step, the generative diffusion model receives a separated token as a condition to predict plausible future motion, and then the selected frames from the predicted motion are utilized as the conditions for the next step generation.
    Unlike other sequential generation methods, which often struggle to quickly adapt to changes in another person's motion, our autoregressive manner can react to the partner's motion effectively, ensuring a more realistic interaction.
    }
    \label{fig:network}
\end{figure*}

\subsection{Reactive Dual-streaming Motion Generation}
\label{sec:multi-condition}

Fig.~\ref{fig:network} illustrates the architecture of our dual-streaming motion generator. 
In an autoregressive manner, the generated motion will be used as the condition for the generation in the next window.
The key of our system is a generative diffusion model $\net$, which takes multiple conditions as inputs to predict a plausible future motion.
More details about the model design and learning strategies are provided below.

\subsubsection{Multi-conditional Generator}
\df{We use a diffusion model \cite{ddpm_ho} as our probablistic generator.}
Given a clean data sample $\sample_0$, the forward diffusion process incrementally adds Gaussian noise to $\sample_0$ by $T$ steps, generating a sequence of noisy samples $\sample_1, ..., \sample_T$: $q_\theta(\sample_t|\sample_0) = \mathcal{N}(\sample_t; \sqrt{\bar{\alpha}_t}\sample_0,(1 - \bar{\alpha}) \mathbf{I})$, where $\bar{\alpha}_t = \prod_{s=1}^t(1-\beta_s)$ and $\beta_t$ denotes the variance scheduler of the added noise.
Then a conditional model $\net$ is trained to learn the reverse process that sequentially recovers the clean sample from random noise over T steps: $p(\sample_{t-1}|\sample_t, c)$, where $c$ is the given condition vector.
The prediction is formed by:
\begin{equation}
		\samplepred_0 = \net(\sample_t, t; c ).
\end{equation}
where the data sample $\sample_0$ and conditions $c$ represent the collection for different types of data, which will be detailed in the following sections. 

\paragraph{Separated conditional tokens}

Our auto-regressive generator iteratively produces the next window of motion based on the \df{following inputs:}
Past self-motion $\mathbf{m}$, which leverages previous movements to ensure the continuity of the generated motion;
Future trajectory $\mathbf{p}$, including the root position and facing direction of the person; 
Future speech features $\mathbf{s}$, as outlined in Section \ref{par:speech}, comprise semantic, melodic, and rhythmic components ($\mathbf{s}^{\text{sem}}$, $\mathbf{s}^{\text{mel}}$, and $\mathbf{s}^{\text{rhythm}}$) from the main person's speech; 
Partner's past motion $\mathbf{m}_\text{refer}$ is also included to generate interactive motion in response to the other person's movement; 
The partner's future speech features $\mathbf{s}_\text{refer}$, which include semantic ($\mathbf{s}^{\text{sem}}_\text{refer}$), melodic ($\mathbf{s}^{\text{mel}}_\text{refer}$), and rhythmic ($\mathbf{s}^{\text{rhythm}}_\text{refer}$) elements from the partner's speech. \tk{Using the future information may sound unrealistic for interactive applications, but indeed humans predict it for smooth communication.~\cite{pickering2007people}.
We will leave a model that predicts such information for future work. 
}

We use Transformer~\cite{transformer} as the backbone of our diffusion model.
Different from previous methods that map various condition into a single latent~\cite{pfnn}, or a single token ~\cite{alexanderson2023listen, pang2023bodyformer, gesturediffuclip}, we get inspiration from ~\cite{camdm} that tokenizes different conditions separately and concatenate them one by one with time step embedding  for the further denoising transformer.
This design enhances the interpretability of the model and allows the model to focus on different aspects of the input data. 
This generation process is:
\begin{equation}
    \begin{aligned}
        \hat{\sample}_0 = \net_{m}(\sample_t, t; [\mathbf{m}, \mathbf{p}, \mathbf{s}, \mathbf{m}_{refer}, \mathbf{s}_{refer}]),
    \end{aligned}
\end{equation}

\subsubsection{Reactive Motion Generation}
For producing realistic motion during conversation, addition to feeding the status of the reference person into the generator, we need strategies to improve the generalization to produce interactive motion. We discuss these below.

\paragraph{Random mask on reference}
\tk{
Using two speaker's speech and motion as input can potentially risk the system to mix-up the speaker and non-speaker, resulting in non-speaker moving in synchronization with the other speaker's voice. To avoid this, we extend the concept of classifier-free guidance (CFG) within our system. During training,  we randomly mask the partner's motion for all training pairs with a probability of 0.15, enabling the model to learn motion generation without relying on the partner's motion.
This scheme also allows us to use other single-person speech/motion dataset for training our system~\cite{liu2022beat, ghorbani2022zeroeggs}.  Integrating such unpaired data into our system can enhance the model's generalization, resulting in more diverse and realistic motion generation. We copy the speaker's motion as the partner's motion; the random masking helps the system to avoid relying on the partner's motion when learning on single-person dataset.
 }
 
In the inference, we apply the following CFG equation: 
\begin{equation}
    \begin{aligned}
        \net(\sample_t, t; c) &= 
        \net_{m}(\sample_t, t; [\mathbf{m}, \mathbf{p}, \mathbf{s}, \mathbf{m}_{refer}=\varnothing, \mathbf{s}_{refer}=\varnothing]), \\
        &+~\gamma \bigl( 
            \net_{m}(\sample_t, t; [\mathbf{m}, \mathbf{p}, \mathbf{s}, \mathbf{m}_{refer}, \mathbf{s}_{refer}]) \\ 
            &- \net_{m}(\sample_t, t; [\mathbf{m}, \mathbf{p}, \mathbf{s}, \mathbf{m}_{refer}=\varnothing, \mathbf{s}_{refer}=\varnothing]) 
        \bigr).
    \end{aligned}
    \end{equation}
It can balance the conditional and unconditional generation, adjusting the influence of the reference motion by the hyper-parameter $\gamma$. This approach offers a more flexible method for generating motion across various scenarios.

\paragraph{Alternant root position normalization}

In single-person motion generation, the first frame of each motion clip is typically shifted to the root. This normalization process is intended to enhance the model's inference generalization. However, applying this method to two person's motion data would lead to the loss of essential relative positional information.
To maintain both model generalization and relative positional information, we randomly designate one person as the primary and the other as the secondary. The primary person's motion is shifted to the root, and the secondary person's motion is adjusted by the same offset. This ensures that the reference motion, \( \mathbf{m}^{\text{refer}} \), retains the relative positional information between the two individuals, allowing the model to learn motion generation in relation to the other person's spatial context.

\subsubsection{Training Loss}
In addition to the MSE loss $\loss_{samp}$ that evaluate the 
local joint rotation difference between the predicted motion and the ground truth motion, we also use the joint position loss $\loss_{pos}$ and velocity loss $\loss_{vel}$ computed by a differentiable forward kinematic layer, and the smoothness loss $\loss_{smo}$ that evaluates the difference of the last frame of the previous motion and the first frame of the future motion. 
The foot contact loss $\loss_{foot}$ is included to ensure the velocity of contact points is close to zero.
The total loss is defined as:
\begin{equation}
    \begin{aligned}
        \loss_{motion} =  \loss_{samp.} + \lambda_{pos}\loss_{pos} + \lambda_{vel} \loss_{vel} + \loss_{smo} + \loss_{foot},
    \end{aligned}
\end{equation} where $\lambda_{pos} = 0.2$ and $\lambda_{vel}=0.5$ are the hyper-parameters to balance the loss terms from different scaling aspects.

\subsection{Autoregressive Framework}
\label{sec:autoregressive}

In the autoregressive system, the motion of two persons is generated simultaneously but in different channels. 
\df{For each person, a sliding window of $N=45$ frames is used, which includes 10 frames from the past and $(N-10)$ frames from the future. The window moves forward by $(N-10)$ frames at a time, causing an overlap of 10 frames. This means the last 10 frames of one window will be included in the next window.}

\paragraph{Stable and Smooth Clip Transition}

Error accumulation is a common issue in autoregressive systems. 
To address this, we introduce a trajectory blending to ensure the stability \df{of long sequence generation}.
In each generation window, the last 10 frames of the previously generated motion are used to predict a trajectory ($\hat{p}^{self}$). This predicted trajectory is then combined with the input trajectory ($p$) using a weighted sum, helping to avoid producing unrealistic motion by reducing the risk of out-of-distribution conditions.
Another issue is the sudden change between the generated clips. To address this, we use a blending strategy known as "dead blending"~\cite{deadblending}. 
\my{This involves considering the final $M$ frames of the preceding clip (source) and the initial $M$ frames of the subsequent clip (destination).}
By smoothly merging the extrapolated segment of the source clip with the commencement of the destination clip, we can produce smooth and continuous motion transitions. 
More runtime strategies are introduced in the supplementary material.

\paragraph{Real-time Performance}

\df{We use an 8-step diffusion model similar to~\cite{camdm} to achieve real-time performance.}
The motion prediction module comprises a 4-layer transformer, equipped with 4 attention heads. The feed-forward dimension and the hidden dimension of the transformer are set to 2048 and 256, respectively.
This configuration allows our model to generate each motion clip within 8ms, facilitating a generation speed of over 100fps, which is well-suited for real-time applications.
Moreover, our clip-based generation strategy gives the flexibility to adjust the frequency of motion prediction. A lower prediction \my{frequency} can reduce the computing cost but may lead to less responsibility, while a higher prediction rate can provide more smoothness but may lead to more latency.

%% file: sec/4_experiments.tex
\section{InterACT++: Extended Co-Speech Interaction Dataset}

\df{InterAct~\cite{arfriend} is the only publicly available dataset for speech-driven two-person interaction.}
It captures conversational dynamics reflective of daily life scenarios, containing 241 two-person motions, and audio sequences with a total of 8.3 hours. 
However, upon detailed manual analysis, we identified a notable limitation in the diversity of interactions. Specifically, the interactions involve less movement, with the participants either stationary or engaging in simplistic walking patterns. Consequently, the dataset lacks sequences of common and dynamic interactions, such as hugging, handshaking, etc.

\df{To address this, we enrich the dataset with a wider array of interaction patterns.}
The enriched dataset includes physical actions such as hugging, handshaking, waving, grabbing the other's hand, patting, and giving high-fives. For each action, 10 scenarios containing plausible situations in which the action may be performed are devised, and each scenario is recorded 4 times for data completeness (twice with actors sitting, twice with actors standing). The newly collected data has a total of 402 clips and 1.7 hours. 
We call the extended dataset as InterAct++.
Table \ref{tab:dataset_comp} shows a comparison between the previous dataset and our new dataset.

\begin{table}[h]
    \caption{Comparison of existing body-centric dataset. 
    InterAct~\cite{arfriend} is the only dataset that contains two-person motion and audio sequences.
    \df{We enrich it with InterAct++ by including 402 clips about dynamic and common two-person interactions.}}
    \resizebox{\columnwidth}{!}{
    \begin{tabular}{c|ccc|cccc}
        \toprule
        dataset & body & face & audio & interaction & \#seq & dura & avg.len \\
        \midrule
        BEAT \cite{liu2022beat} & \checkmark & \checkmark & \checkmark & - & 2508 & 76h & 109s \\
        InterHuman \cite{liang2024intergen} & \checkmark & - & - & \checkmark & 7779 & 6.56h & 3s \\
        ReMoCap \cite{ghosh2024remos} & \checkmark & - & - & \checkmark & 112 & 2.04h & 65s \\
        InterAct \cite{arfriend} & \checkmark & \checkmark & \checkmark & \checkmark & 241 & 8.3h & 60s \\
        Ours (InterAct++) & \checkmark & \checkmark & \checkmark & \checkmark & 402 & 1.7h & 15s \\
        \bottomrule
    \end{tabular}}
    \label{tab:dataset_comp}
\end{table}

\section{Experiments}

\df{Given a two-person conversation, our method can generate contextually meaningful motions of two characters.} %
\df{As existing works either focus on co-speech gesture synthesis of a single person, or text-driven two-person interaction generation, there are no concurrent works that are directly comparable with our method. We thus evaluate both parts separately. In Sec.~\ref{sec:eval_audio2pose} we compare with the co-speech synthesis methods under the two-person setting. Then we evaluate the system on the interaction generation task in Sec.~\ref{sec:eval_inter} by comparing with other interaction generation methods.}
Additional evaluation can be found in the supplementary.

\subsection{Experiment Setup}

We use the original InterAct dataset~\cite{arfriend} and our extension to evaluate the system. The dataset is split into training and test sets with a ratio of 8:2 with their original split. 
\df{The trajectory predictor is trained to only receive audios as input, which aligns to the co-speech gesture synthesis baselines.}
As our method requires 10 frames from the past motion, we simply duplicated the T-pose to fulfill the requirement.
For other datasets we involved like BEAT~\cite{liu2022beat},
as a common practice, we retarget all the motion into the same skeleton with fingers, and downsample the motion from 60 to 30 FPS, and the audio to 16 kHz.

\paragraph{Metrics}
The crux of evaluating the system lies in the quality of the generated motion and its synchronization with the speech. More specifically, we use the following metrics to evaluate the system:
1) Frechet Pose Distance (FPD)~\cite{liu2022beat}: the Frechet Inception Distance between the generated and ground truth pose of each character, evaluating the realism of the generated motion;
2) Frechet Distance-Matrix Distance (FDD). For an interactive motion pair, we calculate the per joint distance matrix and compute the Frechet Distance to evaluate the realism of interactions.
3) Beat Align (BA)~\cite{li2021ai}: the synchronization between the generated motion and the speech, by using the average distance between every kinematic beat and its nearest music beat;
4) Diversity (DIV): the diversity of the generated results, which is computed as the average pairwise L1 distance of all generated clips;
5) Foot Sliding (Foot.Slid): the foot sliding in the generated motion. We use a threshold of 3cm to determine the contact status;
6) Inference Time (Inf.Time):  the average inference time over generating 100 frames of motion.

\subsection{Results}

\begin{figure}[h]
	\centering
	\includegraphics[width=1\columnwidth]{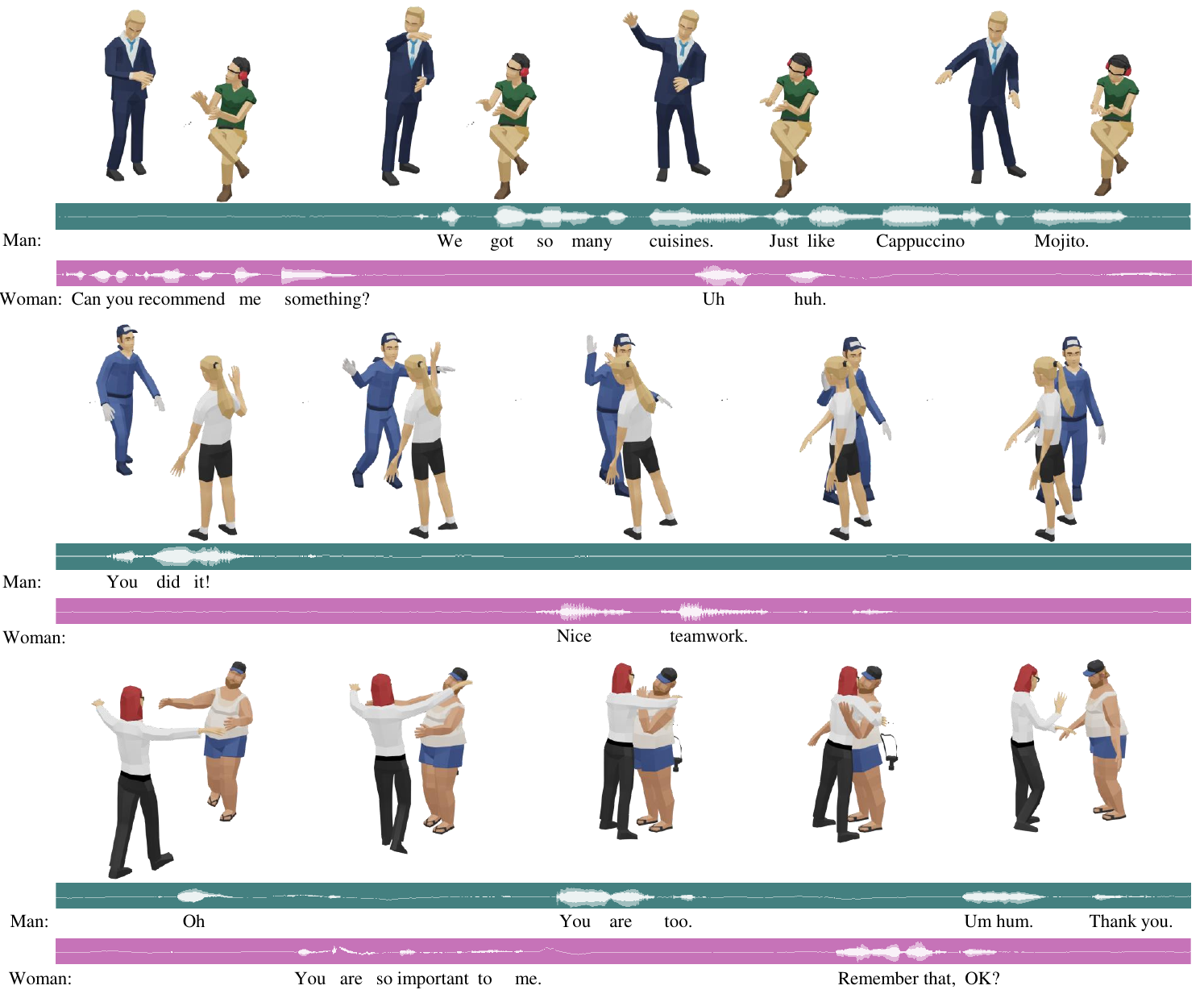}
	\caption{Our generator delivers realistic interaction between two people in sync with the speech.}
	\label{fig:result}
\end{figure}

\begin{figure*}[t]
    \centering
    \includegraphics[width=\textwidth]{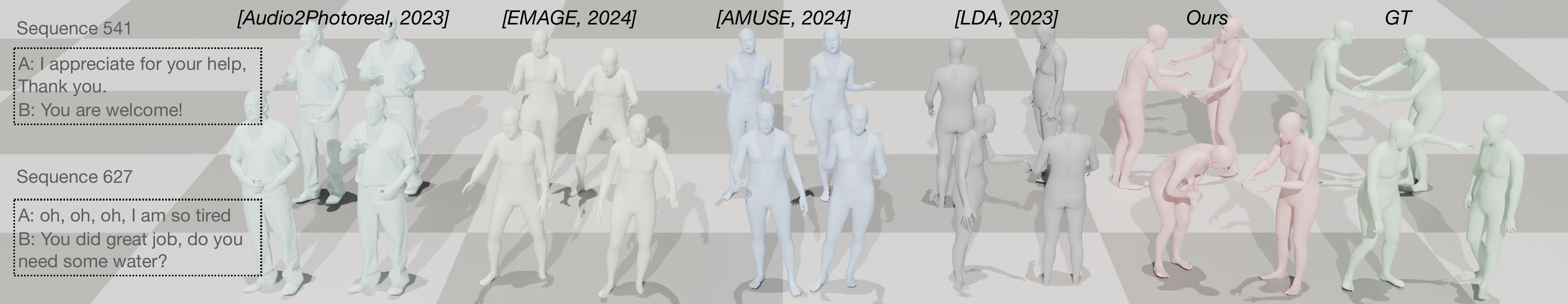}
    \caption{The qualitative comparison among various co-speech methods. We use a consistent SMPL-X representation for the mesh rendering, except for Audio2Photoreal~\cite{li2021audio2gestures} which provides its own mesh template. Single-person co-speech methods, such as Audio2Photoreal, EMAGE~\cite{emage2024} and AMUSE~\cite{Chhatre_2024_CVPR}, fall short in capturing interaction and producing dynamic motion. While the trajectory-based method LDA~\cite{alexanderson2023listen} succeeds in creating dynamic motion, its lack of a reactive generation mechanism hampers realism. 
    }
    \label{fig:result_comp}
\end{figure*}

In Fig.~\ref{fig:result}, we illustrate the capability of our system to generate co-speech two-person interactions. This is evidenced by the depiction of directional movements of crucial body parts such as arms, demonstrating the realistic and fluid interaction between the two individuals. The transcripts showcase the precise alignment between the spoken words and the corresponding motions.
Our approach yields results that are notably more dynamic and interactive in comparison to existing methods, 
such as Audio2Photoreal~\cite{li2021audio2gestures}, EMAGE~\cite{emage2024}, AMUSE~\cite{Chhatre_2024_CVPR} and LDA~\cite{alexanderson2023listen},
(see Fig.~\ref{fig:result_comp}).

\subsection{Comparisons}

\subsubsection{Speech2Motion}
\label{sec:eval_audio2pose}

As a co-speech gesture generation task, we compare our method with other closely related single-person audio-to-motion methods. %
For this comparison, \my{we train all of the methods on the BEAT~\cite{liu2022beat} dataset}. %
Then, for the two-person speech-to-motion task, we purposed a new benchmark and compare our method with the variants of LDA~\cite{alexanderson2023listen} on the InterACT++ dataset.

\begin{table}[h]
    \caption{Comparison on the single person Speech2Motion Task. All the methods are trained on the BEAT dataset unless specified. The top three are feed-forward methods and the bottom two are diffusion methods.}
    \resizebox{\columnwidth}{!}{%
    \begin{tabular}{l|ccccc} 
    \toprule
     Methods & FPD$\downarrow$ & BA$\uparrow$ & Div.$\uparrow$ & Foot.Slid.$\downarrow$ & Inf.Time$\downarrow$ \\
     \midrule
     CaMN~\cite{liu2022beat} & 47.20 & 0.39 & 1.29 & 0.0181 & 7.1ms\\
     \citet{habibie2021learning} & 43.59 & 0.39 & 1.47 & 0.0195 &  14ms\\

     EMAGE~\cite{liu2024emage} & 18.80 & 0.77 & \textbf{5.27} & 0.0068  & 35.4ms\\
     \midrule
    AMUSE~\cite{Chhatre_2024_CVPR} & 34.02 & 0.79 & 1.81 & 0.0054  & 90ms \\
     Ours & \textbf{12.85} & \textbf{0.79} & 4.25 & \textbf{0.0032} & \textbf{4ms} \\
     \bottomrule
    \end{tabular}%
   }
    \label{tab:comp-single}
\end{table}

As illustrated in Table~\ref{tab:comp-single}, our method \df{trained solely on BEAT} \my{demonstrates the state-of-art performance compared to other co-speech generation techniques. 
Our results are more stable, showing the benefits of involving the trajectory as a control signal. 
Thanks to fewer diffusion steps and smaller model size, our method also takes minimum inference time, which is particularly vital for online applications.}

\begin{table}[h]
    \caption{The benchmark on the InterACT++ dataset for two-person speech-to-motion task. All the methods are trained on the train split, and the evaluation is conducted on the test split. LDA-dual is the dual-person version of LDA~\cite{alexanderson2023listen} that takes two-person audio as one input \df{and generates two-person motion together.}}
    \resizebox{\columnwidth}{!}{%
    \begin{tabular}{l|ccccc|ccc} \hline
     & \multicolumn{5}{c}{Motion Quality} & \multicolumn{2}{|c}{Interaction}\\ \hline
     & FPD$\downarrow$ & BA$\uparrow$ & Div.$\uparrow$ & Dyn.$\uparrow$& Foot.Slid.$\downarrow$ & FDD$\downarrow$ & Div.$\uparrow$ \\ \hline
     Audio2Photoreal~\cite{ng2024audio2photoreal} & 130.63 & 0.21 & 3.19 & 1.93 & 0.0382 & 563.27 & 3.77 \\
    LDA~\cite{alexanderson2023listen} & 89.42 & 0.24 & 5.77 & 1.48 & 0.0251 & 583.46 & 3.94 \\
    LDA-dual~\cite{alexanderson2023listen} & 73.20 & 0.33 & 4.29 & 2.56 & 0.0291 & 165.83 & 8.59 \\
    IntetAct~\cite{arfriend} & 55.43 & 0.39 & 9.72 & 2.47 & \textbf{0.0098} & 298.34 & 7.41 \\
    Ours & \textbf{47.74} & \textbf{0.63} & \textbf{14.80} & \textbf{6.54} & 0.0104 & \textbf{117.88} & \textbf{12.54} \\ \hline
    \end{tabular}%
    }
    \label{tab:comp-two}
\end{table}

The benchmark on the two-person speech-to-motion task is shown in Table~\ref{tab:comp-two}. 
Normal single-person co-speech method, such like Audio2Photoreal~\cite{ng2024audio2photoreal} claims they are conversational aware, but it's still challenge without reactive design.  
Even though LDA can generate more dynamic co-speech movements, it still lacks the ability to produce interactive motion. 
Modeling the two-person's motion as a whole, LDA-dual is able to capture the some interactions, but the seq2seq fashion compromises the diversity of the generated motion, also the motion quality is not as good as our method because of weak generalization. 
Our method demonstrated the best performance on both the motion quality and the interaction quality. 
More comparison can be found in supplementary material.

\subsubsection{Interaction Generation}
\label{sec:eval_inter}

For evaluating the effectiveness of our system in generating interactive motion, we deactivate the audio condition and compare our method with other interaction generation methods, such like ReMoS~\cite{ghosh2024remos} which is the state-of-the-art methods that generate interactive motion. 

\begin{table}[h]
    \caption{Comparison on the interactive motion generation task. We take the GT motion of one character as input to generate the motion of the other character.}
    \centering
    \resizebox{0.9\columnwidth}{!}{%
    \begin{tabular}{l|ccc|ccc} \hline
     & \multicolumn{3}{c}{Motion Quality} & \multicolumn{2}{|c}{Interaction}\\ \hline
     & FPD$\downarrow$ &  Div.$\uparrow$ & Foot.Slid.$\downarrow$ & FDD$\downarrow$ & Div.$\uparrow$ \\ \hline
    ReMoS~\cite{ghosh2024remos} & 475.32 & 111.90 & 0.3050 & 394.7 & \textbf{34.0} \\
    Ours & 103.19 & \textbf{10.42} & \textbf{0.0109} & 133.72 & 14.13 \\ 
    \midrule
    Ours(w/t audio) & \textbf{86.79} & 14.84 & 0.0141 & \textbf{104.43} & 19.90 \\ \hline
    \end{tabular}%
    }
    \label{tab:comp-inter}
\end{table}

The results are reported in Table~\ref{tab:comp-inter}. Because of less contacts in our conversational senerio, the explicit constrain purposed in other methods doesn't work well, resulting in worse performance compared to original paper. 
Our method achieves the best motion quality, with the lowest FPD and foot sliding. Even the diversity score of ReMoS is the highest, but based on our observation, it is mostly due to noisy results.
In the audio invloved senerio, it helps to further improve the interaction quality, showing the evidence than audio is an significant factor for generating interaction.

\subsection{Ablation Study}

In our system, we have several modules that are designed to improve the motion quality and realism for the interaction. We conduct an ablation study in Table~\ref{tab:ablation} to evaluate the effect of each module.

\begin{table}[h]
    \caption{Ablation study of different system module.}
    \resizebox{\columnwidth}{!}{%
    \begin{tabular}{l|cccc|ccc} \hline
     & \multicolumn{4}{c}{Motion Quality} & \multicolumn{2}{|c}{Interaction}\\ \hline
     & FPD$\downarrow$ & BA$\uparrow$ & Div.$\uparrow$ & Foot.Slid.$\downarrow$ & FDD$\downarrow$ & Div.$\uparrow$ \\ \hline
    w/o ref.Mot. & 49.22 & 0.61 & 15.25 & 0.0093 & 441.50 & 8.59 \\
    w/o traj. & 103.49 & 0.41 & 10.23 & 0.0297 & 181.95 & 7.42 \\
    w/o SCT & 75.01 & 0.49 & 9.51 & 0.0237 & 204.19 & 6.83 \\
    w/o blending & 54.73 & 0.62 & 14.94 & \textbf{0.0102} & 118.52 & 12.61 \\
    \midrule
    Full & \textbf{47.74} & \textbf{0.63} & \textbf{14.80} & 0.0104 & \textbf{117.88} & \textbf{12.54} \\ \hline
    \end{tabular}%
    }
    \label{tab:ablation}
    \end{table}

Considering the overall quality of the generated interaction, our final system achieves the best balance in producing the results.
The system without high-level speech information and semantic BERT features cannot yield reasonable interaction, and the reference motion is also helpful to generate rhythmic movements.
The model without audio as input, can produce more smooth and high-quality movements, however they are not aligned well to the conditions. 

\subsection{User Study}

The user study was conducted through a questionnaire distributed among participants with diverse backgrounds. The participants were asked their preferences based on three criterias listed below. We provide 20 samples for each, and finally collect 232 users' feedback, in total 4640 data points.
The results, expressed as the percentage of user preferences for each method, are presented in Table \ref{tab:user_study}.

\begin{table}[h]
    \caption{Percentage of user preferences for each method under different criteria.}
    \centering
    \resizebox{\columnwidth}{!}{%
    \begin{tabular}{l|ccc} \hline
    Percentage(\%) & Speech-motion Align & Animation quality & Interaction \\ \hline
    LDA & 13.3 & 23.5 & 4.5 \\
    LDA-dual & 11.5 & 14.1 & 13.0  \\
    Ours & {\bf 75.2} & {\bf 62.4} & {\bf 82.5} \\ \hline
    \end{tabular}%
    }
    \label{tab:user_study}
\end{table}

%% file: sec/5_conclusion.tex
\section{Conclusion}

This work presents a novel method for generating two-person interaction from speech. 
It is the first autoregressive co-speech system which supports real-time generation of full-body motion of two characters from speech, showing the potential into reactive applications.
Moreover, we collect a new dataset which includes more interactions and demonstrates improvements in adopting it into the training. 

\paragraph{Limitations}

Firstly, the absence of an explicit constraint on body joint positions means some interactions, such as handshaking, may not be physically plausible. 
Additionally, capturing data for two individuals is more challenge, particularly in scenarios involving extensive hand interactions. The limited scale of our current dataset may impede the model's ability to generalize and produce responses to arbitrary speech and higher-quality results.
Lastly, our method does not include facial animation, which are also important for expressing realistic interactions.

%% file: sec/X_suppl.tex
\clearpage
\setcounter{page}{1}
\maketitlesupplementary

\appendix

This supplementary material summarizes the video content in~\ref{sec:supmatvid} and provides more information about our data collection process in~\ref{sec:supmatdatacollect}. It further elaborates on our data representation strategies in~\ref{sec:supmatdata} and offers technical specifics about our system, including trajectory prediction, model architecture, training details, and runtime strategy in~\ref{sec:supmatsys}. Finally, we add more details for the evaluation metrics used in~\ref{sec:supmatmetrics}.

\section{Supplementary Video} 
\label{sec:supmatvid}

The supplementary video includes additional results and comparisons to demonstrate the effectiveness of our approach.
Specifically, it provides:
\begin{enumerate}
  \item Two-person's co-speech motion generation results,
  \item Demonstration for the dataset,
  \item Comparisons with AMUSE~\cite{Chhatre_2024_CVPR} and EMAGE~\cite{emage2024}  on single-person's co-speech motion generation,
  \item Comparisons with LDA~\cite{alexanderson2023listen} and Audio2Photoreal~\cite{ng2024audio2photoreal} on two-person conversational generation,
  \item Comparisons with~\cite{ghosh2024remos}  on two-person interaction generation,
  \item Ablation comparisons of the two-person audio conditions and the trajectory conditions.
\end{enumerate}

\section{Data Collection}
\label{sec:supmatdatacollect}

\paragraph{Content}

In InterAct++, 8 actors, consisting of 4 males and 4 females, were recruited for the capture. The actors performed the interactions in male-male or female-female pairs, filling in gender combinations that were lacking in the previous dataset. Table \ref{tab:scenarios} shows excerpts of scenarios included in our extended dataset.

\begin{table}[h]
    \centering
    \begin{tabular}{cc}
        \toprule
        Action & Sentence incl. in scenario \\
        \midrule
        \multirow{3}{*}{Hug}
        & Do you mind giving me a hug? \\
        & I've missed you so much. \\
        & It's so good to see you. \\
        \midrule
        \multirow{3}{*}{Handshake}
        & Pleased to meet you. \\
        & Hope we can collaborate again soon. \\
        & Thank you for the opportunity. \\
        \midrule
        \multirow{3}{*}{Wave}
        & Catch you next time! \\
        & Looking forward to seeing you again! \\
        & Have a great day! \\
        \midrule
        \multirow{3}{*}{Grabbing hand}
        & Careful, you almost fell. \\
        & You're not going anywhere! \\
        & Hang on, I've got you. \\
        \midrule
        \multirow{3}{*}{Pat}
        & There, there. It's going to be okay. \\
        & Don't worry about it, you'll get it next time. \\
        & Proud of you. \\
        \midrule
        \multirow{3}{*}{High-five}
        & That was awesome! \\
        & Teamwork makes the dream work! \\
        & Victory! \\
        \bottomrule
    \end{tabular}
    \caption{Excerpt of scenarios in our capturing.}
    \label{tab:scenarios}
\end{table}

\paragraph{Capture Process}

\begin{figure}[h]
    \centering
    \includegraphics[width=1\columnwidth]{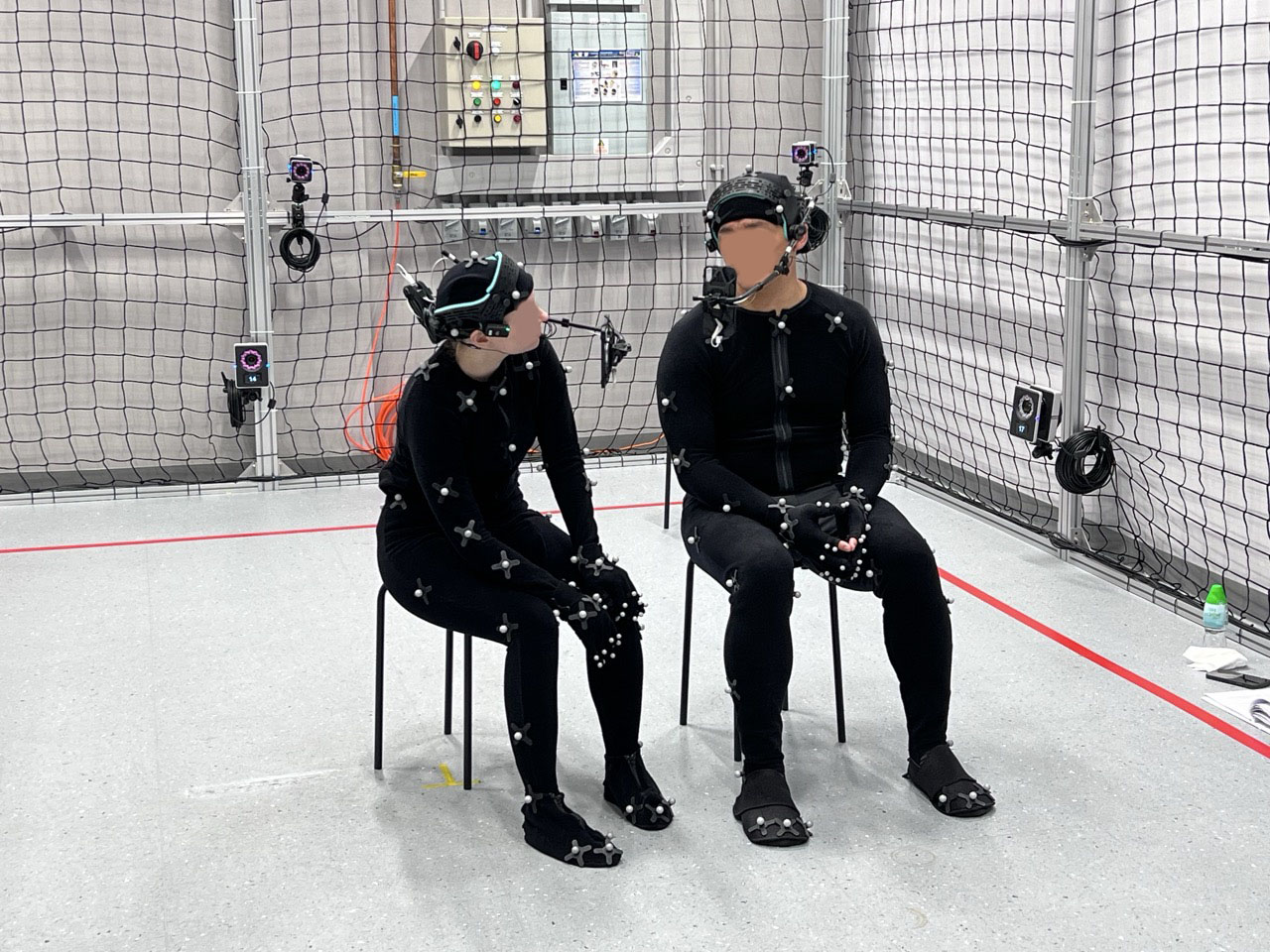}
    \caption{A pair of actors wearing full capture outfits, recording audio, facial expressions, and body movements.}
    \label{fig:capture_actors}
\end{figure}

The capture system comprises various components, either worn by the actors or positioned throughout the capture area. To monitor the body movements of both actors, a 28-camera VICON optical motion capture (MoCap) system is arranged around a 5m x 5m space, with cameras set at three different heights. Each actor is outfitted with 53 body markers and 20 finger markers, and they perform a Range of Motion exercise for calibration prior to the capture. An iPhone's front depth sensor and camera are utilized to record each actor's facial expressions. Before capturing, a mesh template of each actor's face is created by having them rotate their head while facing the camera. During the capture process, an actor wears a head rig that holds an iPhone, two microphones (one primary and one backup), and a power bank that also serves as a counterweight. Figure \ref{fig:capture_actors} displays a snapshot of two pairs of actors during data collection.

Since the motion and facial capture systems operate independently, they must be synchronized temporally to ensure accuracy. A wireless timecode generator transmits the clock signal from the VICON system to each iPhone, aligning the timecode at the start of capture. A script is executed before and after each capture session to simultaneously initiate and halt recordings for both systems at a specified timecode. This process guarantees frame-level precision while minimizing the need for post-processing to synchronize the two data sources. During the capture, live audio and video footage of both actors are streamed to the control station for real-time monitoring, which helps maintain data quality. If any errors are detected, the capture is paused and then resumed.

\section{Detailed Data Processing}
\label{sec:supmatdata}

\paragraph{Speech Tokenization}

\begin{figure}[h]
	\centering
	\includegraphics[width=1\columnwidth]{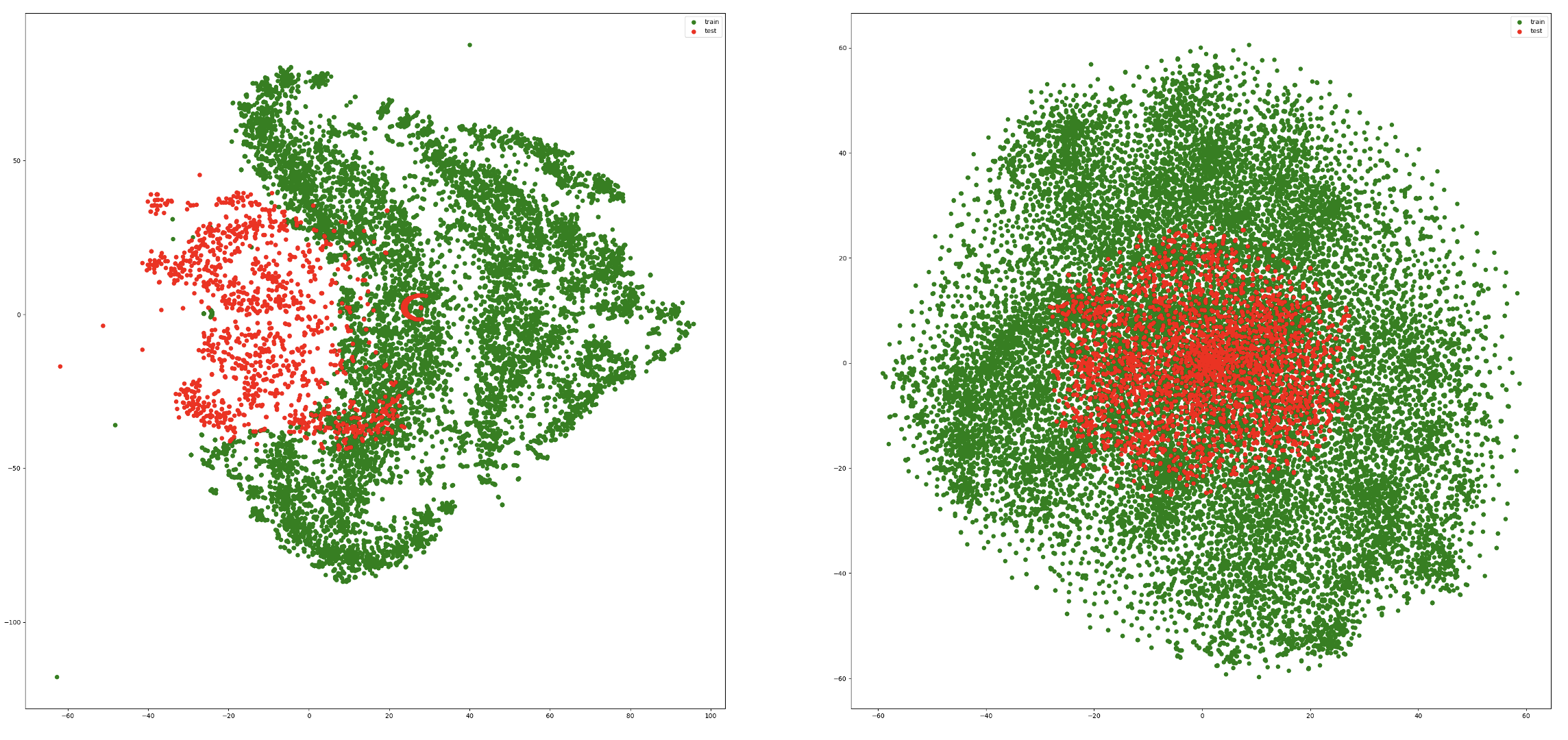}
	\caption{We extract the speech semantic features besed on BERT(Left) and speech tokenization(Right) and then visualize the data points of the \textcolor{green}{training} and \textcolor{red}{testing} data in the 2D space using t-SNE. The distributions are aligned better in the speech tokenization space.}
	\label{fig:supptoken}
\end{figure}

As introduced in the main paper, we didn't use the text-based embedding to represent the semantic information of the speech because of the overcrowding problem. Instead, we used speech tokenization~\cite{zhang2023speechtokenizer} which is trained on the large-scale  speech data.
We prove it by visualizing data points of BERT features and discrete speech tokens from our InterAct dataset in Figure~\ref{fig:supptoken}. 
With BERT features, the entire embedding space is shared by a wide array of text-based data. This leads to a reduced convergence probability based on the limited speech transcript from the training split, exacerbating the Out of Distribution (OOD) issue in this task. As shown in Figure~\ref{fig:supptoken} \textit{left}, the testing data are not covered by the training distribution.
In contrast, using discrete speech tokenization effectively mitigates the overcrowding problem. It is trained specifically on speech data, resulting in a more compact representation and enhancing the likelihood of finding alignments between training and testing data. As shown in Figure~\ref{fig:supptoken} \textit{right}, the training audio token distribution fully covers the test data, leading to better performance on inference.

We take the checkpoint called \textit{speechtokenizer\_snake} from the SpeechTokenizer~\cite{zhang2023speechtokenizer} model, which is trained on the English speech datasets including LibriSpeech~\cite{panayotov2015librispeech} and common voice. Each 20ms segment of audio will be quantized into tokens with 8 dimensions, where the first 1 dimension is the semantic information and the rest 7 dimensions are the timbre information. We use the first dimension as the semantic information and ignore the rest 7 dimensions.

\paragraph{Motion Processing}

Except for the newly captured dataset (InterAct++), we involve the BEAT~\cite{liu2022beat}, InterACT~\cite{arfriend} in our experiments.
They all provide the kinematic motion structure, but the skeletons are not the same. 
To reduce the risk of skeleton misalignment, we retarget all the motion data to the Mixamo skeleton structure. It includes 65 joints, including the fingers.
Moreover, we correspond this skeleton with an SMPL-X model for further visualization and comparison.

\begin{figure}[h]
	\centering
	\includegraphics[width=1\columnwidth]{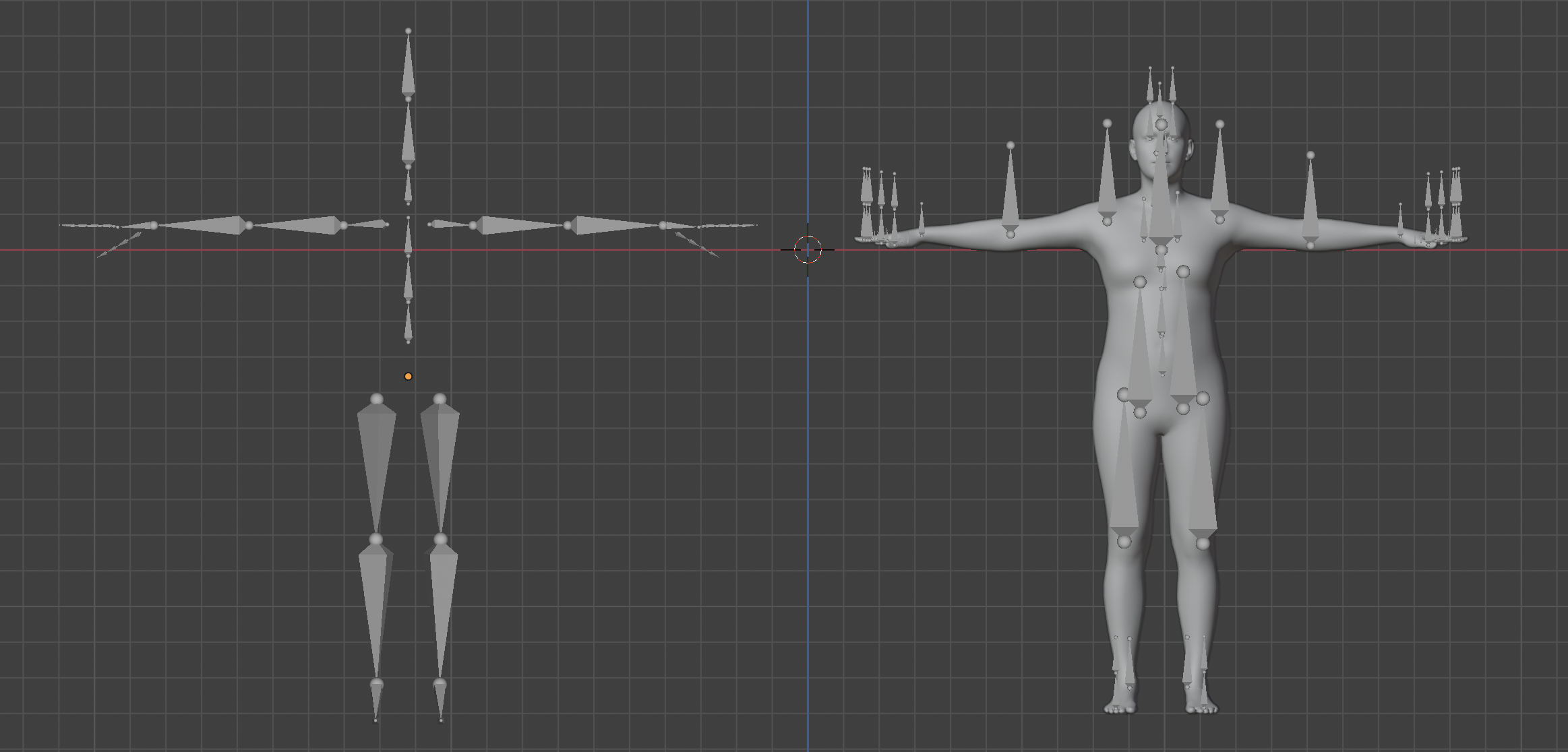}
	\caption{Left: The unified skeleton in our experiments. Right: The aligned SMPL-X model for the visualization. We use our kinematic data to animate the SMPL-X model rather than training with it.}
	\label{fig:suppskeleton}
\end{figure}

\section{System Technical Details}
\label{sec:supmatsys}

In this section, we provide more details about the trajectory prediction module, model architecture, training details, and runtime strategy to help the readers understand our system better.

\subsection{Trajectory Prediction}

Trajectory, a common control signal in animation systems, does not only offer a more stable expression space compared to the original speech, but can also describe the spatial-temporal relations of the characters in a compact fashion. In our experiments, it's significant to help the motion generator to produce stable and controllable motion. Please see the supplementary video for an ablation.

In practice, it can be produced by different methods in specific scenarios. For example, if we just want the person walk forward to a specific position, the trajectory can be generated by the target position and the initial position with a linear interpolation. 
In interactive generation, the user might want to control the two person's interaction via a specific trajectory.

In our system, a more powerful trajectory prediction module is needed because we take into account the face direction of the characters, which is crucial for realistic interaction.
Moreover, we want to generate the trajectory of two persons based on the speech and other global context information, such as how much distance the two persons have moved, and the start/target position of the two persons to achieve high-level control.
For keeping everything in real-time, We use similar auto-regressive model as the motion generation module to predict the trajectory of two persons.

\begin{figure}[h]
	\centering
	\includegraphics[width=1\columnwidth]{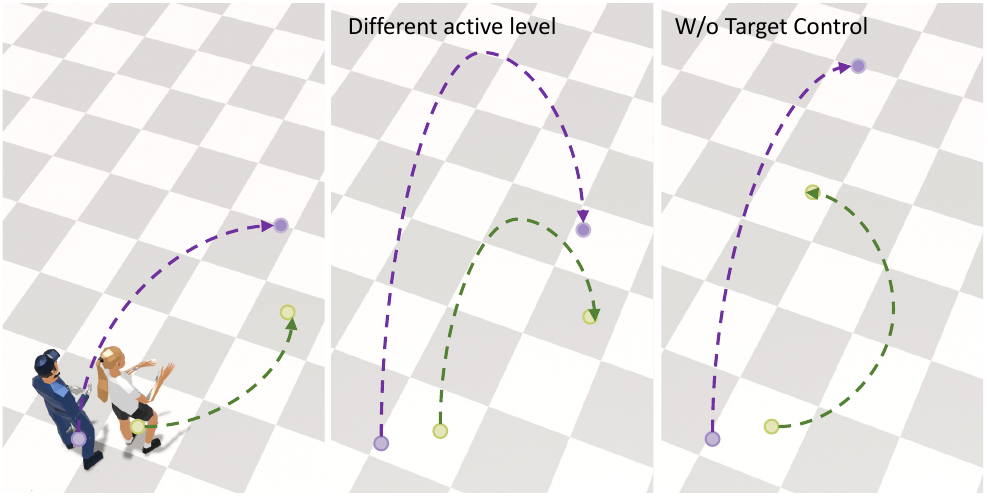}
	\caption{Our trajectory prediction system takes multiple conditions as input, such as speech, high-level activity values, and positional signals. \dfc{the figure is not clear. what are green and purple? the text in the figure is hard to see.}}
	\label{fig:traj}
\end{figure}

\paragraph{Inputs}

As this module focuses on global context to produce two-person's long-term control signals, it takes the following inputs for each person together:
\begin{itemize}
    \item[--] Initial body position and facing direction. 
    \item[--] Speech features $\mathbf{s}^{\text{BERT}}_n$ and $\mathbf{s}^{\text{mel}}_n$. The beat information has less relationship with the trajectory in our experiments, hence we do not use $\mathbf{s}^{\text{rhythm}}_n$ in this module.
    \item[--] Global activity level $\mathbf{f}$,  ranging from 0 to 10. We use the statistical moving distance of the two persons and normalize it as a scalar representing the activity level. %
    \item[--] Target position of the characters. In some cases, the animator may prefer to specify the goal configuration of the two characters. This input can be masked if not needed.
\end{itemize}

These conditions are tokenized separately and fed into the generator for predicting the trajectories. In our experiments, this setting is helpful to produce realistic interactions between characters in response to the context of the speech.  

\paragraph{Training Loss} 
For modeling the relationship between two persons, we generate the trajectory of two persons together and apply a joint MSE loss to the predicted trajectories, which is defined as:

\begin{equation}
    \begin{aligned}
        \loss_{traj} = 
        &\mse(\mathbf{P}^{A}, \net_{traj}(\mathcal{N},[\mathbf{p}^{A}_0, \mathbf{S}^{\text{BERT}}, \mathbf{S}^{\text{mel}}, \mathbf{f}, \varnothing \times \mathbf{p}^{A}_N], t))  \\
        + &\mse(\mathbf{P}^{B}, \net_{\text{traj}}(\mathcal{N},[\mathbf{p}^{B}_0, \mathbf{S}^{\text{BERT}}, \mathbf{S}^{\text{mel}}, \mathbf{f}, \varnothing \times \mathbf{p}^{B}_N], t)).
    \end{aligned}
\end{equation}

During the inference stage, the initial statuses of two individuals can be inferred, randomly decided within a region or set to a default position; the model will direct the characters to a similar trajectory given the same speech input with the same input noise.

\subsection{Model Architecture}

We list the detailed architecture of our model in the following. The model consists of a motion diffusion module, which is a transformer-based model that takes the speech features, trajectory, and other conditions as input and generates the motion of two persons. The model is trained in an autoregressive manner, where the generated motion is used as the input for the next frame generation. 

\begin{lstlisting}[breaklines=true,basicstyle=\tiny]
MotionDiffuison(
  (future_motion_process): MotionProcess(
    (poseEmbedding): Linear(in_features=396, out_features=512, bias=True)
  )
  (past_motion_process): MotionProcess(
    (poseEmbedding): Linear(in_features=396, out_features=512, bias=True)
  )
  (partner_motion_process): MotionProcess(
    (poseEmbedding): Linear(in_features=396, out_features=512, bias=True)
  )
  (audio_process): TrajProcess(
    (poseEmbedding): Linear(in_features=27, out_features=512, bias=True)
  )
  (audio_strength_process): TrajProcess(
    (poseEmbedding): Linear(in_features=1, out_features=512, bias=True)
  )
  (token_process): TrajProcess(
    (poseEmbedding): Linear(in_features=1, out_features=512, bias=True)
  )
  (partner_token_process): TrajProcess(
    (poseEmbedding): Linear(in_features=1, out_features=512, bias=True)
  )
  (sequence_pos_encoder): PositionalEncoding(
    (dropout): Dropout(p=0.2, inplace=False)
  )
  (seqTransEncoder): TransformerEncoder(
    (layers): ModuleList(
      (0-3): 4 x TransformerEncoderLayer(
        (self_attn): MultiheadAttention(
          (out_proj): NonDynamicallyQuantizableLinear(in_features=512, out_features=512, bias=True)
        )
        (linear1): Linear(in_features=512, out_features=1024, bias=True)
        (dropout): Dropout(p=0.2, inplace=False)
        (linear2): Linear(in_features=1024, out_features=512, bias=True)
        (norm1): LayerNorm((512,), eps=1e-05, elementwise_affine=True)
        (norm2): LayerNorm((512,), eps=1e-05, elementwise_affine=True)
        (dropout1): Dropout(p=0.2, inplace=False)
        (dropout2): Dropout(p=0.2, inplace=False)
      )
    )
  )
  (embed_timestep): TimestepEmbedder(
    (sequence_pos_encoder): PositionalEncoding(
      (dropout): Dropout(p=0.2, inplace=False)
    )
    (time_embed): Sequential(
      (0): Linear(in_features=512, out_features=512, bias=True)
      (1): SiLU()
      (2): Linear(in_features=512, out_features=512, bias=True)
    )
  )
  (output_process): OutputProcess(
    (poseFinal): Linear(in_features=512, out_features=396, bias=True)
  )
)
\end{lstlisting}

\subsection{Training Details}

\begin{figure}[h]
  \centering
  \includegraphics[width=1\columnwidth]{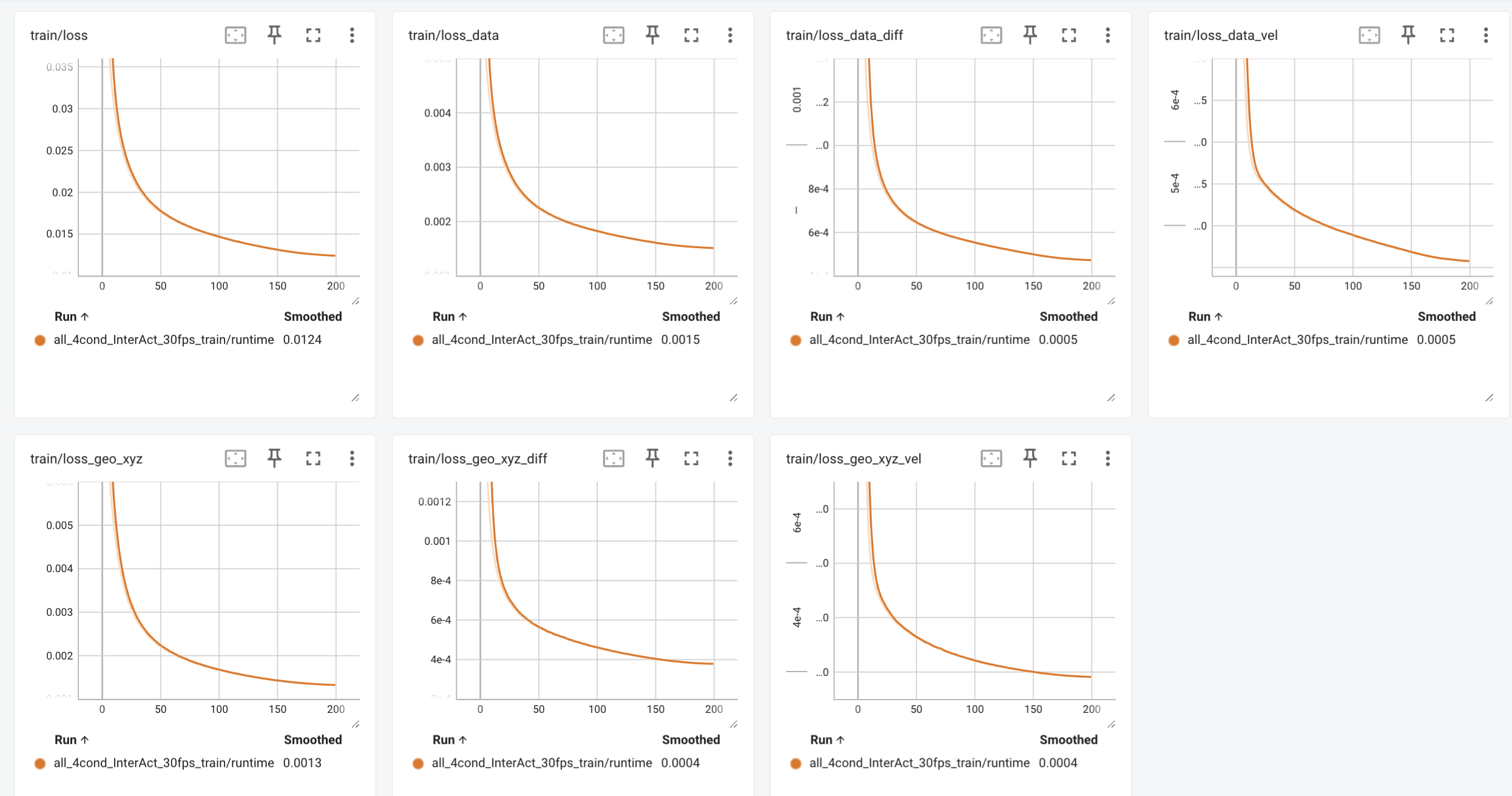}
  \caption{The training loss and validation loss of our model.}
  \label{fig:supploss}
\end{figure}

In our experiments, the entire training process is conducted on a single NVIDIA A100 GPU. The model is trained with a batch size of 1024 and a learning rate of 0.0005. The model is trained for 200 epochs, taking around 2 days in our InterACT++ dataset. 
Longer training is beneficial for the model because of our runtime augmentation strategy, which can generate motion with different face directions.
We visualize the training loss and validation loss in Figure~\ref{fig:supploss}.

\subsection{Runtime Strategy}

We use a sliding window to generate the motion, where the last frames of the previous clip are used as the control signal for the next clip generation. This strategy reduces the computational cost and ensures the stability of the generated motion.
The windows size and applied frame numbers can be adjusted according to the scenario. 
In our experiments, we use 10 past frames to predict the next 45 frames of motion. Then we only taks the first 15 frames and play it, the rest 30 frames will be used as the control signal for the next clip generation as introduce in next subsection.

\paragraph{Trajectory Blending}

The trajectory blending is a common way in autoregressive motion generation~\cite{pfnn, localphase} to ensure the stability and controllability of system.
In each sliding generation window, the last 10 frames of generation motion can provide a self-predicted trajectory $\hat{p}^{self}$, which can be used as the control signal for the next clip generation.
By default, we use the extension function $HFTE$ introduced by ~\cite{camdm} to extend the predicted trajectory of the same frame as the control signal, which operates by sequentially repeating in the local frame curve flip, denoted as $HFTE(\hat{p}^{\text{self}})$.
We then blend the input given trajectory ${p}$ with the self-predicted trajectory by using a weighted sum:
\begin{equation}
    \begin{aligned}
        \hat{P}^{\text{blend}} = {p} \times \alpha + HFTE(\hat{p}^{\text{self}}) \times (1-\alpha),
    \end{aligned}
\end{equation}
where the $\alpha$ is a scalar value that controls the influence of the input trajectory on the generated motion. %
The amount of $\alpha$ could be adjusted according to the scenario; for example, when the two person are moving to a specific position, a bigger $\alpha$ will be more suitable.

\paragraph{Motion Clip Transition}

Our approach generates motion in a sequential, clip-by-clip manner, with each subsequent clip overlapping the input window. To ensure seamless continuity between transition frames, we utilize a blending strategy known as "dead blending"~\cite{deadblending}. This involves considering the final $M$ frames of the preceding clip (source) and the initial $M$ frames of the subsequent clip (destination). We extend the duration of the source animation beyond the transition point. Subsequently, we apply a traditional cross-fade blending technique. This method involves smoothly merging the extrapolated segment of the source clip with the commencement of the destination clip.
This method is simple and effective to produce smooth and continuous motion transitions.

\section{Evaluation}
\label{sec:supmatmetrics}

\subsection{Metrics}

For the FPD, we refer to ~\cite{glide} that uses the pose skeleton to calculate the distance between the generated motion and the ground truth.
For the FDD, we also compare the disbution on the joint distance matrix level.
The dynamic metric is measure by the average moving distance of the two persons in the generated motion.

\subsection{LDA}

LDA ~\cite{alexanderson2023listen} is a method that receives the audio as input to generate single-person motion. And we adopt it in following design:
1) LDA-single, we trained the original LDA model with same dataset to produce the motion for each person separately; 
2) LDA-dual, which use the audio of both persons as input to generate the motion for each person separately.